# A new 2D monolayer BiXene, M2C (M=Mo, Tc, Os)

Weiwei Sun[a],[✉], Yunguo Li[b],[†], Baotian Wang[c,d], Xue Jiang[e], Mikhail I. Katsnelson[f,g], Pavel Korzhavyi[h,i], Olle Eriksson[a], Igor Di Marco[a],[‡]

aDepartment of Physics and Astronomy, Material theory, Uppsala University, Box 516, SE-75120 Uppsala, Sweden; ✉sun.weiwei@physics.uu.se; ‡Corresponding author: Igor Di Marco, igor.dimarco@physics.uu.se

bDepartment of Earth Sciences, University College London, London WC1E 6BT, UK; †yunguo.li@ucl.ac.uk

cChina Spallation Neutron Source (CSNS), Institute of High Energy Physics (IHEP), Chinese Academy of Sciences (CAS), Dongguan 523803, China

dDongguan Institute of Neutron Science (DINS), Dongguan 523808, Chinaa

eKey Laboratory of Materials Modification by Laser, Ion and Electron Beams (Dalian University of Technology), Ministry of Education, Dalian 116024, China

f Radboud University of Nijmegen, Institute for Molecules and Materials, Heijendaalseweg135, 6525 AJ Nijmegen, The Netherlands

gTheoretical Physics and Applied Mathematics Department, Ural Federal University, Mira Street 19, 620002 Ekaterinburg, Russia

hDepartment of Material Science and Engineering, KTH-Royal Institute of Technology, Stockholm SE-10044, Sweden;

iInstitute of Metal Physics, Ural Division of the Russian Academy of Sciences, 620219 Ekaterinburg, Russia

The existence of BiXenes, a new family of 2D monolayers is here predicted. Theoretically, BiXenes have 1H symmetry ($P\bar{6}m2$) and can be formed from the $4d/5d$ binary carbides. As the name suggests, they are close relatives of MXenes, which instead have 1T symmetry ($P\bar{3}m1$). The newly found BiXenes, as well as some new MXenes, are shown to have formation energies close to that of germanene, which suggests that these materials should be possible to be synthesised. Among them, we illustrate that 1H-$Tc_2C$ and 1T-$Mo_2C$ are dynamically stable at 0 K, while 1H-$Mo_2C$, 1T-$Tc_2C$, red1H-$Tc_2C$, and 1T-$Rh_2C$ are likely to be stabilised via strain or temperature. In addition, the nature of the chemical bonding is analysed, emphasizing that the covalency between the transition metal ions and carbon is much stronger in BiXenes than in MXenes. The emergence of BiXenes can not only open up a new era of conducting 2D monolayers, but also provide good candidates for carrier materials aimed at energy storage and spintronic devices that have already been unveiled in MXenes.

**Introduction**

The emergence of graphene has promoted the discovery and investigation of novel two-dimensional (2D) materials [1,2], as for instance elementary silicene and germanene [3,4], functional graphene, binary single layer boron-nitride [5], as well as monolayer transition metal dichalcogenides [6]. Their fascinating and exotic electronic, optical, mechanical, magnetic and thermal properties provide massive possibilities to synthesise artificial materials for potential applications in catalyst, new energy areas, *et cetera* [7]. Due to these high expectations, the 2D materials world is constantly enriched by the discovery of more and more emerging materials [8]. Nevertheless very few conductors with pronounced metallic character have been found so far, which is a limit for technological applications. Among the best conductors one can find MXene films, which are a class of 2D early transition metal carbides (MCs) and carbonitrides with 1T symmetry [9]. The metallic conductivity of MXenes is also accompanied by surface hydrophilicity, which means that they behave as "conductive clays" [10]. Hence, MXenes have attracted more and more attention, because they have a tremendous potential for applications in electrode materials, sensors, catalysis and electrochemical energy storage [11–14].

MXenes are produced by immersing selected MAX phase powders in hydrofluoric acid (HF). MAX phases include compounds of composition $M_{n+1}AX_n$, where M is a transition metal element, A is an element mostly from IIIA and IVA columns, and X is carbon or nitrogen. The index n can be 1, 2 or 3, depending on the geometrical arrangement of atoms [15,16]. In principle, many MCs can be intercalated by light ions to form 2D materials, providing several fascinating properties like superconductivity, good electrochemical performance and storage properties for hydrogen and lithium ions [13]. Unfortunately, up to date, existing 2D materials derived from MCs are limited to MXenes, and involve mostly 3$d$ M and a very limited number of heavier elements. [17]

We recently reported [18] that at ambient conditions bulk $Ru_2C$ can be stable in the space groups $R3m$ and $R\bar{3}m$. These structures are composed of layers that are ideal candidates as novel 2D systems. It is natural, therefore, to explore the structural and dynamical stability of 2D monolayers arising from $Ru_2C$ and neighbouring systems, $Tc_2C$, $Rh_2C$ and $Os_2C$, which are also likely to exist in a layered structured. To these compounds, one can add $Mo_2C$, which was recently studied in a couple of experiments. First, a large surface of superconducting $Mo_2C$ thin film of 10 Å thickness was synthesised at a high temperature [19]. The preparation procedure was based on a modification of the method used for growing high-quality graphene, chemically-functionalized and defective reduced graphene oxide, or also for preparing MXenes. This approach may be used as a general strategy for fabricating monolayer carbides and other 2D crystals, including transition metal nitrides (MNs) [20]. Further, the large-scale synthesis of 2D $Mo_2C/Mo_2CT_x$ flakes (where T is a surface termination group, like e.g. -O -OH or -F) was also realized, via selective etching of Ga from $Mo_2Ga_2C$ powders [17,21]. Holding to these clues, it becomes crucial to understand the structural and dynamical stability of 2D monolayers arising from Mo, Tc, Ru, Rh, and Os binary carbides.

We hereby predict the existence of a new family of 2D materials having 1H symmetry and involving two layers of heavy M atoms sandwiching a graphene-like interlayer of C atoms. Due to the fact that they originate from binary carbides and have the same chemical composition of MXenes, we name these systems BiXenes. BiXenes differ from MXenes not only for their symmetry, but also for the nature of the M components. While MXenes are mostly composed of light M elements, e.g. 3$d$ elements, BiXenes contain $4d/5d$ elements, which results in wider $d$ bands. This is likely to bring out different functionalities like magnetic or superconducting properties [22]. As for several known 2D monolayers, distinctive geometry and compositions are of great importance for improving mechanical stability and related properties [23]. The emergence of BiXenes offer a truly exciting opportunity that can lead to dozens of new metallic 2D materials, possibly showing exotic properties.

**Computational methods**

In this work we performed *ab initio* electronic structure calculations based on density-functional theory (DFT). We used the Vienna ab-initio Simulation Package (VASP) [24,25], based on a projected-augmented wave (PAW) method [26]. Within PAW, 4p, 4d, and 5s states were treated as valence states for Mo, Tc, Ru, and Rh; 5p, 5d, and 6s states were used for Os and 2s and 2p states for C. The Perdew-Burke-Ernzerhof (PBE) exchange-correlation functional in the generalized-gradient approximation (GGA) [27] was used for all elements. An energy cutoff of 650 eV was employed to achieve the energetic convergence of 0.001 meV as well as the force convergence of 1 meV/Å. The formation energies of the bulk were evaluated using the diamond structure as a reference state for C. Using diamond instead of graphite is computationally easier since one does not need to include Van der Waals (VdW) interactions. The reference state of the M elements was instead obtained from their corresponding ground-state structure, i.e. bcc for Mo, hcp for Tc, Ru and Os, fcc for Rh. A more detailed analys of the motivations and consequences of these choices can be found in our previous work on $Ru_2C$ [18]. Finally, we sampled the Brillouin Zone (BZ) with a dense Monkhorst-Pack k-grid [28] of 17 17 5 points.

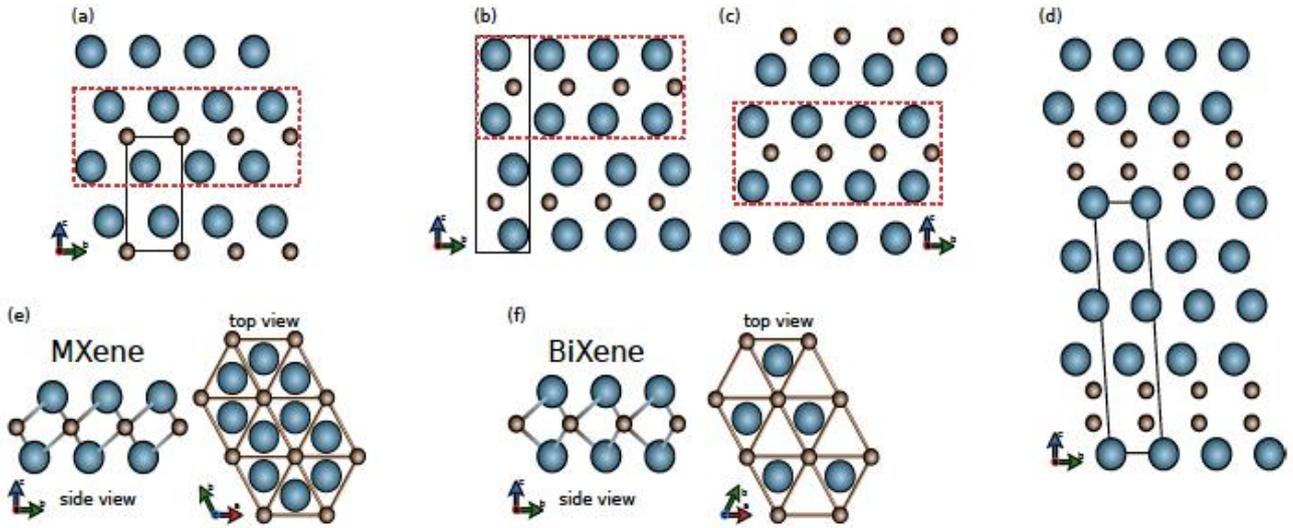

**Fig. 1** Crystal structures for the 3D and 2D materials studied in this work. For 3D materials: (a) $P\bar{3}m1$; (b) $P6_3/mmc$; (c) $R3m$; (d) $R\bar{3}m$. For 2D materials: (e) MXene 1T ($P\bar{3}m1$); (f) BiXene 1H ($P\bar{6}m2$). The blue spheres stand for M atoms and the brown spheres stand for C atoms. The dotted red rectangles in panels (a)-(d) represent the periodic block composing the 2D materials in panels (e)-(f). The black lines identify the unit cell.

For the 2D materials, a supercell was constructed including a vacuum layer with a thickness of at least 15 Å. Full optimisation was carried out in the in-plane directions. The formation energies $E_{form}$ with respect to the bulk were evaluated as:

$$E_{form} = E_{2D}(M_2C) \quad E_{bulk}(n[M_2C])/n \quad (1)$$

where $E_{2D}$ and $E_{bulk}$ are the total energies of 2D monolayer and bulk, respectively, while $n$ stands for the number of formula units in bulk unit cell. The energies of the bulk were calculated with respect to the $P\bar{3}m1$ for MXene, and the energetically favoured structure between $R3m$ and $P6_3/mmc$ for BiXene (see also Fig. 1). For sake of completeness, for the 2D materials, another set of calculations was performed including VdW interactions by means of the optB86b-VdW functional [29,30]. The BZ was sampled with a dense Monkhorst-Pack k-grid [28] of 17 17 1 points. Note that for these open structures, an atomic centred basis set is much more precise. Therefore, we analysed the Fermi surfaces (FS) by means of the full-potential linear muffin-tin orbital (FP-LMTO) code by M. Van Schilfgaarde et al. [31].

The density functional perturbation theory (DFPT) implemented in the PHONOPY code [32] was employed to examine the dynamical stability. To ensure a reasonable convergence, we used an energy cutoff of 700 eV with a precision of $1 \times 10^{-8}$ eV. The force constants in the bulk were calculated for an isotropic 5 5 5 supercell whose BZ was sampled with 5 5 5 k-points. Those in the 2D materials were instead obtained for an isotropic 5 5 1 supercell whose BZ was sampled with 5 5 1 k-points. For $Mo_2C$, it was necessary to extend the 2D supercell to 6 6 1. The Fermi-Dirac (FD) broadening scheme [33] for an electronic temperature s was used in the self-consistency cycle to smear out the abrupt change of the Fermi-Dirac statistics in the ground state.

### Results and discussion

**From 3D to 2D $M_2Cs$** We first investigate the bulk $M_2Cs$ in the $P\bar{3}m1$, $P6_3/mmc$, $R3m$ and $R\bar{3}m$ structures (see panels (a)-(d) of Fig. 1). In the SI we provide an overview of ground-state structures of the bulk, formation energies and phonon spectra, as well as motivations to restrict our analysis to the four aforementioned structures.

The 3D structures are composed of layers of different symmetries as basic building blocks. As shown in panel (e), in the $P\bar{3}m1$ bulk structure one can identify layers with 1T symmetry with inversion (point group $D_{3d}$). This corresponds to the 2D structure of MXenes, mentioned above. As shown in panel (f), instead, from the $P6_3/mmc$ and $R3m$ bulk structures, one can extract 2D systems with the 1H symmetry without inversion (point group $D_{3h}$). This corresponds to the 2D structure of $MoS_2$ but has never been found before for carbides or nitrides [8]. Moreover, there is a fundamental feature which makes BiXene different from the transition metal dichalcogenides $MX_2$ (X=S, Se, Te). This is that metal and non-metal atoms are swapped. The different arrangement leads to very different electronic structures and related properties, as we will see in section . Finally, it is worth stressing that both the 1H and 1T structures exhibit hexagonal lattices, while their main geometrical difference concerns the M layers, which are overlapping in the top view in BiXene (see Fig. 1).

The relative energies shown in the upper panel in Table 1 indicate that either 1H or 1T is more energetically favoured. In the framework of PBE, all 1H-$M_2$Cs are favoured, with the exception of 1T-$Rh_2$C. Including VdW interactions makes also 1T-$Os_2$C more favourable than the correspONding 1H structure. We notice that for the early elements Tc and Mo, results are only barely affected by including VdW interactions, while for the other systems the corrections are larger. We are not sure that this trend is physical, since it may be related to the particular VdW implementation used in VASP, whose v[34].

**Table 1** Top panel: relative energies (in eV) of the 2D monolayers (0 eV corresponds to the most stable structure). Bottom panel: formation energies of the 2D monolayers with respect to the bulk, as from Eq. (1). The energies are given in eV per formula unit, for both the 1H and 1T structures, referred to their most energetically favourable 3D counterpart.

|  | PBE | | | | | VDW | | | | |
|---|---|---|---|---|---|---|---|---|---|---|
|  | $Mo_2C$ | $Tc_2C$ | $Ru_2C$ | $Rh_2C$ | $Os_2C$ | $Mo_2C$ | $Tc_2C$ | $Ru_2C$ | $Rh_2C$ | $Os_2C$ |
| $E_{Rtot}$ (1H) | 0 | 0 | 0 | 0.25 | 0 | 0 | 0 | 0 | 0.24 | 0.01 |
| $E_{Rtot}$ (1T) | 0.19 | 0.25 | 0.08 | 0 | 0.11 | 0.24 | 0.27 | 0.49 | 0 | 0 |
| $E_{form}$ (1H) | 2.06 | 2.82 | 2.19 | 1.58 | 2.57 | 2.39 | 2.91 | 2.57 | 1.90 | 2.92 |
| $E_{form}$ (1T) | 2.69 | 2.17 | 1.81 | 1.38 | 1.81 | 3.03 | 2.52 | 2.56 | 1.66 | 2.06 |

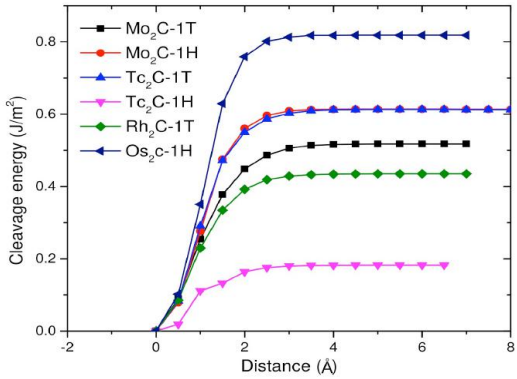

**Fig. 2** (Colour online) Cleavage energy $E_{cl}$ in J/m2 as a function of the separation distance d for a fracture in bulk materials for the six (quasi) stable bixenes.

In addition to the direct comparison between 3D and 2D materials, the cleavage energies are also considered as it is actually a measurement.

Most of 2D materials are exfoliated slices of powders (bulk)/thin films or can be prepared by designed methods as e.g. chemical vapour deposition (CVD)[19]. Their feasibility can at first be evaluated through their formation energies with respect to the bulk, i.e. $E_{form}$ from Eq. (1). The values for the 2D materials addressed in this work are reported in the lower panel of Table 1. In general, most of the formation energies appearing in Table 1 are smaller than the values of the primary (intra-layer) chemical bonds, suggesting that the interaction between the M layers is sufficiently weak for possible exfoliation. In fact, we emphasise that, if normalised per atom, the values reported in Table 1 are comparable to (or smaller than) the formation energies of germanene and silicene[38]. This trend is confirmed by the inclusion of VdW interactions. As expected, the latter make the bulk energetically more convenient, increasing the formation energies of the monolayers. Despite this small increase, the formation energies are still below 1 eV per atom. Overall, all these 2D compounds seem possible to be realised, provided that they are dynamically stable (see below). $Mo_2C$ stands out as the only material for which the 1H structure seems easier to realise than the 1T structure, with a difference in the formation energies of about 0.6 eV. Finally, an interesting speculation based on Table 1 concerns the dependence of $E_{form}$ on the atomic number. For M-ions heavier than Tc, the 1T structure seems more favourable, while the opposite holds for lighter elements. The Tc ion in $Tc_2C$, corresponding to a half-filled 4d-shell, seems in fact to act as a transition element, although more compounds should be investigated to support this analysis.

**Dynamical stability and vibrational properties**

Realising the monolayers in experiments requires also the dynamical stability with respect to the lattice vibrations. This can be analysed by means of the phonon spectra, which are reported in the left panel of Fig. 3. In general, one cannot identify a trend where 1H or 1T polymorph seems to have stable 2D compounds. Moreover, one can notice a general softening for compounds in the 1H structure, which is also seen in the 2D M-dichalcogenides[39,40]. Among all investigated systems, only 1T-$Mo_2C$ and 1H-$Tc_2C$ satisfy the strictest criterion for dynamical stability, i.e. the absence of imaginary branches. Nevertheless, 1T-$Tc_2C$, 1T-$Rh_2C$, 1H-$Mo_2C$ and 1H-$Os_2C$ are characterized by imaginary modes of very small size, which makes them interesting for our purposes. In fact, these small pockets of imaginary modes may be the result of numerical truncation in our theory. In the method (PHONOPY+VASP) employed in this work, due to the missing of transition from the real space to the reciprocal space in the perturbation theory, a supercell is still needed. The numerical truncation arises from the finite size of these supercells, and also from other minor computational details.

Furthermore one has to consider that in experimental conditions 2D materials are usually grown on top of a substrate, such as Si or $SiO_2$. In this theoretical work simulations were performed for the free-standing 2D materials, but experimentally these 2D materials are grown on top of a substrate that provides a (limited) strain. This strain can then be used to realise an otherwise dynamically unstable 2D material. For example, MXenes are usually prepared via chemical etching from the bulk MAX phases, and the monolayers would shrink after the drying in the air[42], but also the TMD, e.g. $WS_2$, and few layers $TaS_2$ monolayers are often grown on SiO2 substrates[43,44]. Therefore, the compressive strain should naturally be applied to the theoretical models.

On the other hand, thermal effects would bring some changes in the electronic system as well as the lattice (phonons). The electronic system is affected through the repopulation of the electronic states by means of the Fermi-Dirac distribution and also through more complex thermal fluctuations of pure many-body nature. The former are easy to describe in DFT, while the latter require a temperature-dependent exchange-correlation functional and are therefore out of reach for our analysis [45]. The stabilisation of the phonon spectra induced by a finite temperature through the Fermi-Dirac smearing is mainly related to the changes induced on the electronic density by the repopulation of the Kohn-Sham orbitals. The precise mechanism is therefore strongly system-dependent, since it depends on the distribution of the electronic states above and below the Fermi energy. What is more relevant to our discussion is that thermal induced stabilisation is often discussed in literature, e.g. for Al hydrides [46] or Nb/Mo nitrides [47,48]. Therefore, we performed this analysis in our manuscript and compared to effective smearing to the one used in those references(see further discussion concerning Fig. 4). In addition to the electronic effects, the phonon spectra are also directly affected by the temperature in the form of phonon-phonon interactions. These are however not included in our calculations, since they are beyond the harmonic approximation. Anharmonic effects are often responsible of stabilising some phases at high temperature [see Refs. [49–52] at the end of this section]. Including anharmonic effects in our study is however too demanding and requires more sophisticated methods that are not accessible to us, at the moment. The induced strains and the effects due to a finite temperature can increase the stability of a system and remove the small pockets of imaginary modes. Therefore, one can consider 6 (quasi) stable monolayers, including 1T-$Mo_2C$, 1T-$Tc_2C$ and 1T-$Rh_2C$ for the MXene structure, and 1H-$Mo_2C$, 1H-$Tc_2C$ and 1H-$Os_2C$ for the BiXene structure.

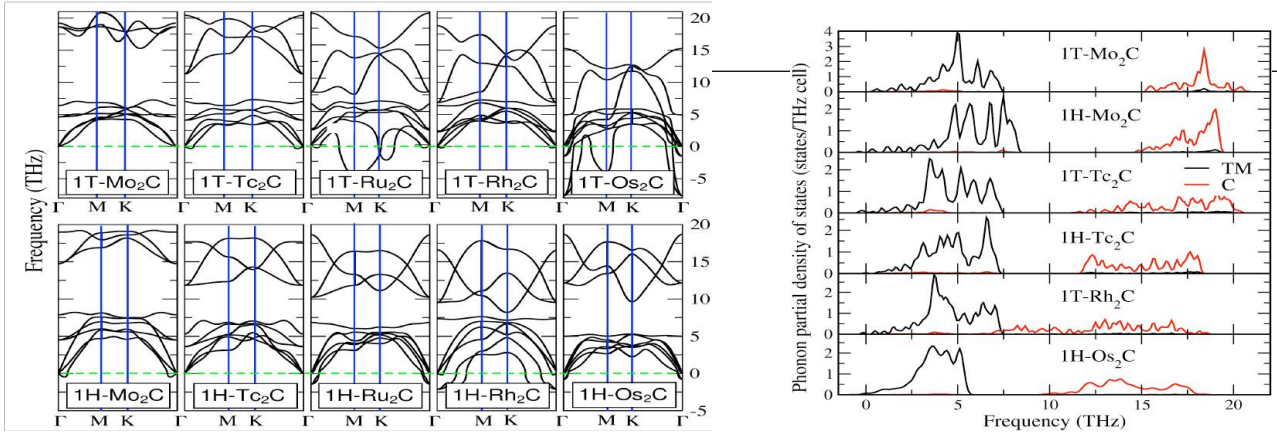

**Fig. 3** (Colour online) Left panel: the phonon dispersion curves for the single layer of $M_2C$ M=Mo, Tc, Ru, Rh, Os with both 1T and 1H symmetries. Right panel: the phonon site-projected density of states (PSPDOS) of 1T-$Mo_2C$, 1H-$Mo_2C$, 1T-$Tc_2C$, 1H-$Tc_2C$, and 1T-$Rh_2C$ and 1H-$Os_2C$ (from top to bottom).

The vibrational properties of these 6 structures are discussed by means of the phonon site-projected density of state (PSPDOS), reported in the right panel of Fig. 3. A general feature for all materials but 1T-$Rh_2C$ is the presence of a big gap between modes projected over M and C species, which is related to their large mass ratio. For 1T-$Rh_2C$, the unusually wide phonon band of the C atoms extends from 18 THz to 8 THz, where it overlaps with the M band. The absence of a gap will be further discussed in the next subsection. In the low frequency region of all materials one can also notice an overlap of M and C modes, corresponding to co-vibrations. This overlap is weak but noticeable, especially for 1T-$Tc_2C$, at around 4 THz. Concerning the general differences between the 1T and 1H structures (to be inferred from both $Mo_2C$ and $Tc_2C$), they appear to be not too marked. Once can notice that in 1H a marked peak characterises the top of the M band, and also that the end of the C band is not as high in energy as in 1T. Finally, we note that the phonon spectrum of $Os_2C$ is slightly different from the other ones, as the M band is narrower and no multiple peaks can be resolved. This may be a consequence of the fact that Os is a 5$d$ element, i.e. much heavier than the 4$d$ elements involved in the other materials. However, it may also be connected to a chemical bonding of a different nature, as arising from the more delocalised 5$d$ electrons. This issue will be addressed more in detail in the next subsection.

As we stressed previously, the quasi stable structures can in principle be stabilised by the presence of a substrate or by effects due to a finite temperature. In perspective of practical applications, it is important to analyse these factors more in detail. Among the quasi stable structures, 1H-$Mo_2C$ is the most interesting system to address, since it offers the lowest formation energy compatible with the 1H symmetry (see Table 1). Moreover, 1T-$Mo_2C$ has been already identified as a very promising candidate for a high-performance thermoelectric material, and its function-alized 2D layer could find many applications within power generation, energy storage devices, and catalysis [53,54]. This suggests that its counterpart 1H-$Mo_2C$ may have equally interesting properties. Therefore, we choose 1H-$Mo_2C$ as a test example to investigate the effects of strain and temperature on the dynamical stability.

**Fig. 4** The phonon dispersion curves of the 1H-$Mo_2C$ monolayer in LDA and under different conditions: (a) free strain, zero temperature; (b) biaxial compressive strain of 3% in the basal plane; (d) biaxial tensile strain of 3% in the basal plane; (d) biaxial tensile strain of 3% in the basal plane at an effective temperature (Fermi-Dirac smearing) of s =0.4 eV. (e-f) the phonon dispersion curves calculated in the small displacement approach [41], (e) in LDA by using the same structure optimized in LDA, (f) in PBE by using the same structure optimised in PBE.

In Fig. 4 we report phonon spectra of 1H-$Mo_2C$ under various conditions. Given that GGA often tends to underestimate the binding, the exchange-correlation functional was switched from GGA to LDA for these calculations. The phonon spectrum with no strain at zero temperature, shown in panel (a) of Fig. 4, is very similar to the corresponding spectrum of Fig. 3. As expected, the imaginary modes are slightly enhanced and also the upper phonon branches (mainly due to C) are upshifted, revealing that LDA strengthens the Mo-C bonds. These effects are further emphasized if a biaxial compressive strain (3%) is applied, as illustrated in panel (b) of Fig. 4. Instead, applying a tensile strain (3%), leads to strong downwards shift of the upper phonon branches, as illustrated in panel (c) of Fig. 4. Moreover, the imaginary modes are strongly enhanced with respect to the unstrained system. Let us now consider the worst case, namely the phonon spectrum of 1H-$Mo_2C$ under a 3% tensile strain, and include temperature effects in the form of a Fermi-Dirac smearing, as usually done for this type of studies [55]. The panel (d) of Fig. 4 shows results for an effective (smearing) temperature of s = 0.4 eV, and we note that there is no trace of negative phonon branches. As a term of comparison we report Fig. 4 (e-f), the phonon spectrum obtained by using the small displacement method (frozen phonons) as implemented in the Phon code [41]. The imaginary modes are still present but strongly quenched with respect to the calculations performed via DFPT. This comparison is useful to have an idea of the numerical errors associated to the truncation. Although s = 0.4 eV may seem a high value, as it roughly corresponds to T = 4500K, it is in line with values obtained for other systems that have been synthesised. For example, the monolayer and few layers of 2H and 1T-$TaSe_2$ can be stabilised only if s is equal or bigger than 0.68 eV [55]. For the less unstable cases, corresponding to panels (a) or (b) of Fig. 4, the imaginary modes are likely to be stabilised with s in the range of 0.1 or 0.2 eV.

As we mentioned previously, including Fermi-Dirac smearing is the simplest way of considering finite temperature effects that should correspond to a more realistic modelling of experimental conditions. It is however difficult to make a quantitative comparison between temperatures in experiment and temperatures in the Fermi-Dirac distribution, due to the fact that we are not using a temperature dependent exchange-correlation functional. Moreover, even if we were using a complete description of the electronic system, we would also need to include temperature effects for the lattice, i.e. anharmonic effects due to the phonon-phonon interaction [see e.g. Ref. [49] of the main manuscript and Ref. [46–48]]. In this context, it is not easy to determine to what extent the electronic temperature is high with respect to experimental conditions. As for the applied strain, strain on 2D materials has recently been investigated in a variety of works, like e.g. by Z. Guo *et al.* [56] or G. Gao *et al.* [57] In this last work the authors focus on the range of applied strain from -5% to +5%, and we decided to follow a similar criterion, albeit restricted to +/-3%. In Fig. 4(e-f), the phonon spectra calculated in the small displacement method is nearly dynamically stable, so that a rather small applied strain or thermal smearing can remove all the imaginary modes.

Finally, in experimental perspective, one should keep in mind that the temperature discussed here does not correspond to the real temperature, as standard exchange-correlation functionals are able to capture only the "weak" temperature dependence, through the Fermi-Dirac distribution. To obtain a more realistic value of the temperature necessary to reach dynamical stability one should apply more sophisticated methods, such as e.g. the self-consistent *ab initio* lattice dynamics method (SCAILD)[49]. This approach takes into account proper anharmonic effects, providing more realistic estimates of the experimental temperature [49–52].

**Chemical bonding and electronic structure** The next point of analysis concerns the chemical bonding and the electronic structure of the identified (quasi) stable monolayers. This will also help explaining the differences observed in the vibrational properties. In the bottom panel of Fig. 5 the electronic density of states (DOS) of the investigated structures is reported. When going from Mo to Tc and Rh, the inclusion of more and more *d* electrons pushes the Fermi level $E_f$ upwards, as shown by the green dashed lines. In contrast to the bulk, no clear formation of "valley states" dividing bonding from antibonding states [58,59] is observed, except for 1H-$Mo_2C$. This is a very interesting feature associated to the lower dimensionality and arises from quantum confinement effects (periodicity broken along the c axis). Most importantly, all the investigated systems have a strong metallic character, which is a rather rare feature for 2D materials. For example, among the transition metal dichalcogenides $MX_2$, the tellurides show a semi-metallic character, while sulfides and selenides are semiconductors [60,61].

The two structures of $Tc_2C$ show the highest number of states at $E_f$, which is related to that four bands cross the Fermi energy in this case, while there are only three for the other materials (see also next subsection). Another major feature noticeable in the DOS is that the states around $E_f$ have mainly M character, and there is no significant hybridization with the C derived states. The most relevant hybridized M-C states lay instead at high binding energy, several eV below the Fermi energy. Moreover, for 1T-$Rh_2C$ and 1H-$Os_2C$, no pronounced antibonding peak can be observed, due to the shift of $E_f$ induced by filling the *d* shell. In 1H-$Os_2C$, hybridized Os-C states spread over a wider range, which is partially due to the broader 5*d* band with respect to the 4*d* band. This is also noticeable through the high energy tail of the DOS, which in 1H-$Os_2C$ goes down to -8.5 eV, i.e. about 1 eV more than for the other systems. Moreover, although it is not visible from the bottom panel of Fig. 5, the tail does not suddenly stop at -8.5 eV but slowly decays to higher binding energies. This does not happen for the other systems. Finally, there seem to be not obvious differences between the DOS in the 1T and 1H structures, apart that the latter appears to be characterised by slightly broader peaks.

Another interesting feature noticeable from the bottom panel of Fig. 5 is that occasionally a pseudo-gap is formed at about -4 eV. In one case, i.e. for 1T-$Mo_2C$, this pseudo-gap becomes a real gap in the spectrum. This transition from a pseudo-gap to a full gap is related to the formation of unique inter-layered electrostatic interactions, and can be better analysed through the electronic lo-

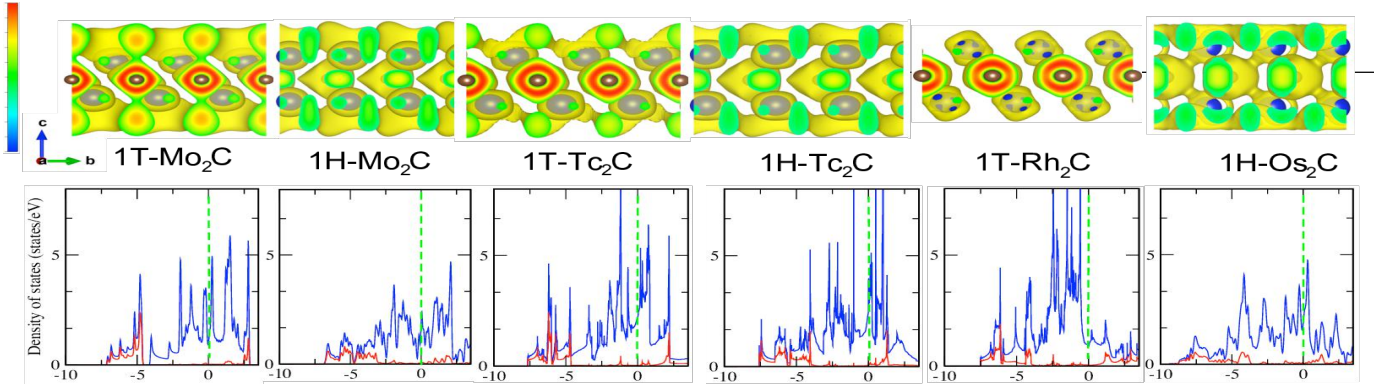

**Fig. 5** Top panel: the electron localisation function (ELF) of the possible (quasi) stable 2D materials: 1T-$Mo_2C$, 1H-$Mo_2C$, 1T-$Tc_2C$, 1H-$Tc_2C$, 1T-$Rh_2C$, and 1H-$Os_2C$. The level of the isosurface is set to be 0.38 of the local maximum value. The blue (grey) spheres represent the M (C). Bottom panel: projected density of states of the M-$d$ (in blue) and C-$2p$ (in red) electrons. The dashed green line indicates the Fermi energy.

2d. calisation function (ELF), shown in the top panel of Fig. 5.

Before discussing the ELF of 1T-$Mo_2C$, we first notice that the ELF of the 1H structures shows a chemical bonding featured with some covalency. The orientations of the ellipsoids of the electron clouds of the C layer are noticeably different. In 1H-$Os_2C$ they prefer to align parallel to the C axis, while in 1H-$Mo_2C$ and 1H-$Tc_2C$ the favoured direction is perpendicular to the C axis. Although the overlap of the electron clouds belonging to the M and C layers can be observed in all three cases, it is particularly strong for 1H-$Os_2C$, where *bridge-like* channels emerge. It can also be inferred that the in-plane C-C p bonds are weaker in the case of 1H-$Os_2C$, and this is different from the other 2D materials.

The 1T structures show conversely more localised charges around the C atoms, which points to that they are more ionic than their 1H counterparts. The ELF of 1T-$Mo_2C$ stands out as significantly different from the others, due to the presence of the point charges at the surface (M layers). These point charges produce a unique inter-layered electrostatic interaction with the C layers, which induce the formation of the gap at -4 eV observed in the bottom panel of Fig. 5. Another interesting feature is that the ELF of 1T-$Rh_2C$ shows a strongly ionic Rh-C bond, where the states at the surface are almost empty. In general, the stronger the inter-layer bonds, the higher the frequency of the C modes in the phonon spectra. Therefore, the ELF data explain the extension of the C bands in the phonon spectra of Fig. 3.

**Fermi surface and band structure**

After having analysed the basic features of the electronic structure and the nature of the chemical bonding, we move to investigate the band structure and the Fermi surface (FS). These are shown in Fig. 6, for the 6 (quasi) stable structures identified above. The band structure of $Tc_2C$, for both symmetries, stands out as it is the only material where four bands cross the Fermi energy. In the FS of 1H-$Tc_2C$ the 1st band forms the small hole pockets, the 2nd band forms the "dog-bone" sheets, the 3rd band forms holes, and the 4th band forms the inner smaller sheets inside the sheet formed by 3rd band. For 1T-$Tc_2C$, the components of FS are very similar to the 1H structure but the "dog-bone" sheets are replaced by a flower-like sheet, as shown in Fig. 6. As expected, filling the 4$d$ shell, as in 1T-$Rh_2C$ or 1H-$Os_2C$, leads to a simpler FS. The hole pockets become smaller, and even disappear in the case of 1T-$Rh_2C$. The most noticeable features are very closely parallel sheets formed by the 1st and 2nd bands, which are evident for both 1T-$Rh_2C$ or 1H-$Os_2C$. In the case of 1T-$Rh_2C$ parallel sheets can be found, whose nesting vector **q** has been defined in Fig. 6. This suggests that a Kohn anomaly [62] may potentially occur at this vector **q**, e.g. in the phonon spectrum, as it happens for e.g. graphene [63] or silicene [64]. Graphene and silicene are in fact the prototypes of 2D monolayers exhibiting Kohn anomalies, which also lead to the formation of charge density waves [65]. Hajiyev *et al.* [66] have also revealed the existence of an incommensurate charge density wave phase transition in single- and double-layered 2H-TaSe, identified by a sudden softening of a Raman mode resulting from the Kohn anomaly through strong electron-phonon coupling. Anyway, one should keep in mind that all these systems are insulating, and can therefore be very different for the present monolayers. Since this nesting direction is not along the high symmetry points, no anomalies have been observed in Fig. 3. Instead, 1H-$Os_2C$ differs more significantly from the other systems, as the FS sheets constructed by 3rd band are no longer two-dimensional due to the complexity of the wiggled red bands shown in Fig. 6. This complexity is likely to arise the fact that each 5$d$ orbital in Os has 12 lobes which show much complexity than that for 3$d$ orbital.

We can then analyse what happens by removing an electron from the half-filled 4$d$ shell. This situation corresponds to a transition from $Tc_2C$ to $Mo_2C$. In 1T-$Mo_2C$ we note that the 1st band forms the overlapping "saddles", while the 2nd band forms hexagons with the holes from the 3rd band. These features are consistent with a simple evolution of the FS of 1T-$Tc_2C$, which has one electron more. Instead, the evolution of the two 1H structures is not so straightforward. In the FS of 1H-$Mo_2C$, there is one less band than in 1H-$Tc_2C$, and the "dog-bone" sheets evolve into smaller holes. Moreover, for 1H-$Mo_2C$, in contrast to 1H-$Tc_2C$, the FS sheets formed by the 2nd band shrink, and additionally there is one less band missing in the FS, also mentioned above. Finally, notice that only 1H-$Tc_2C$ has both hole and electron pockets (see the 1st and 2nd bands in the corresponding electronic structure of

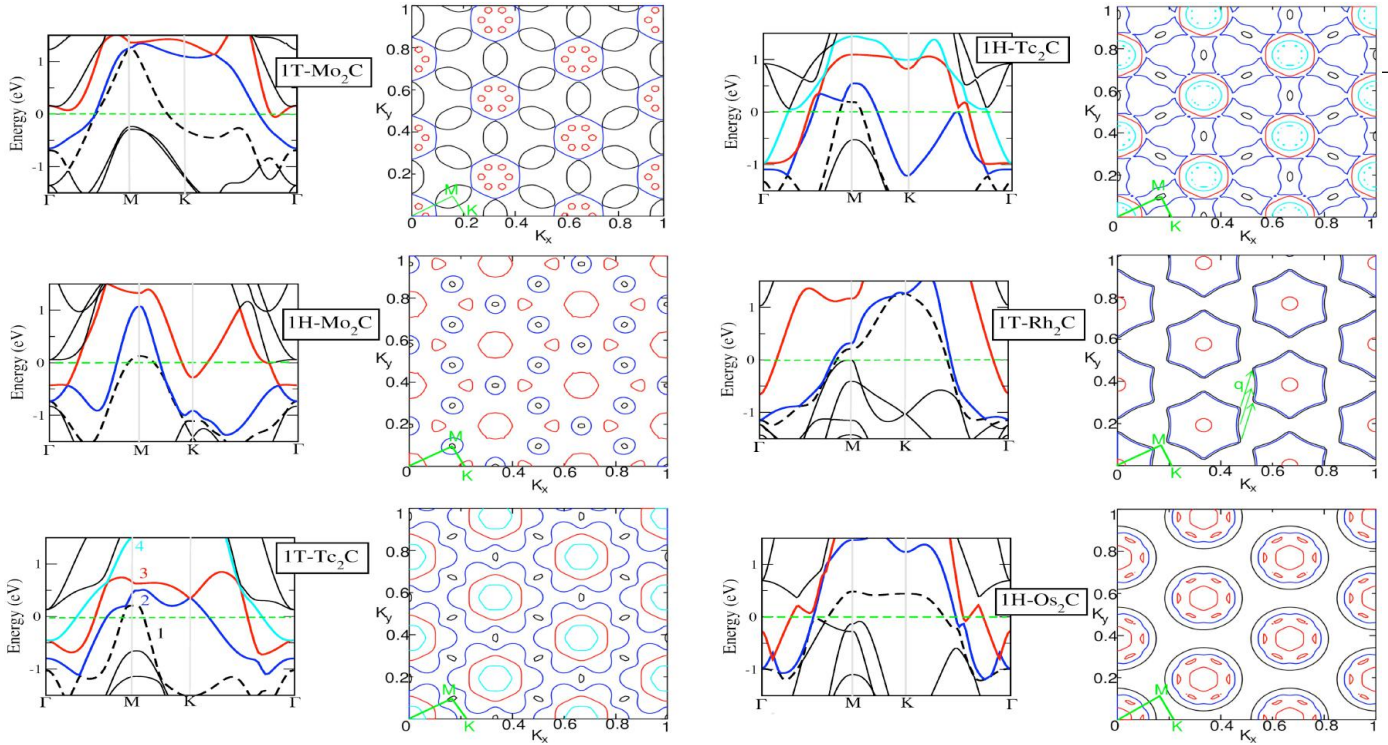

**Fig. 6** Left side: band structure along high symmetry directions for the (quasi) stable 2D materials 1T-$Mo_2C$, 1H-$Mo_2C$, 1T-$Tc_2C$, 1H-$Tc_2C$, 1T-$Rh_2C$, and 1H-$Os_2C$. The bands crossing $E_f$ are labelled by counting their energetic order at the G point. The first band is depicted with dashed black lines, the second one with blue lines, the third one with red lines, the fourth one (when present) with cyan lines. Right side: corresponding Fermi surface (FS) in the plane of $k_x$-$k_y$ in the reciprocal space. The unitary length along both directions corresponds to $3\frac{2p}{a}$. The bands are labelled as in the band structures. The high symmetry points G, M and K, and the paths between them, are shown in green.

Fig. 6).

**Conclusions and Outlook**

A new class of conductive transition metal carbide monolayer BiXene is here theoretically predicted. The BiXene family is reminiscent of the well-known MXenes, but is characterized by 1H symmetry, instead of 1T. The investigation of both BiXenes and MXenes arising from transition metal carbides $TM_2C$, where TM = Mo, Tc, Ru, Rh, and Os, is motivated by their particularly layered structure, composed of a basic building block made by a sandwich of transition metal bilayer along with a graphene-like C layer in the middle. Among the systems studied here, we identify 1H-$Mo_2C$, 1H-$Tc_2C$ and 1H-$Os_2C$ as possible BiXenes to synthesize in future experiments. These materials seem likely to exist, due to relatively low formation energies and stable phonon dispersion curves. The stronger inter-layered interactions of BiXene provide more stable 2D monolayers than MXene, which can be suitable for manipulating carriers in applications and devices. One can also expect larger and more dramatic effects induced by surface modifications and adsorption, which will be the object of our future studies. Also, the Fermi surface nesting found in 1T-$Rh_2C$ opens a link with possible formation of charge-density waves. This, as well as exploring possible structural phase transitions induced by strain or temperature, will be addressed in future studies. Briefly, the emergence of BiXene will stimulate extensive studies on the 2D metallic monolayers, and also supply with another candidate material for nanoscale devices.

In addition to the discovery of BiXene, in this work we also identify three systems in the MXene family, i.e. 1T-$Tc_2C$, 1T-$Rh_2C$, and 1T-$Mo_2C$, which was also recently synthesised[17,21]. These systems enrich the small group of MXenes originating from heavy transition metal carbides and are likely to lead to several interesting properties, which only future theoretical and experimental analyses will unveil. Moreover, the discovery of new pathways to MXenes through binary MCs or MNs will be very useful to overcome the limitation that most MXenes so far required to be produced from MAX phases including Al as A element, which made many predicted monolayers not accessible experimentally[17].

Finally, the new $M_2C$ monolayers are analogous to $MX_2$ in terms of structures (1T and 1H), but differ from the arrangement of the metal and non-metal sites, which is likely to lead to different chemical and physical properties. Our future research will focus on exploring the formation of BiXenes from other carbides and nitrides, possibly exploring all the transition metal elements that can form MAX phases. Furthermore, we will investigate more in detail the role of the temperature in stabilising the group of quasi-stable $M_2C$ monolayers, by means of more sophisticated techniques to describe thermally induced anharmonic effects[49–52].

# 1 Acknowledgments

The simulations were performed on resources provided by the Swedish National Infrastructure for Computing (SNIC) at the High Performance Computing Center North (HPC2N) and at the National Supercomputer Cluster (NSC) in Sweden. The financial support from Knut and Alice Wallenberg Foundation (KAW), Swedish

Research Council (VR), and eSSENCE is acknowledged. The authors appreciate the fruitful discussion with Prof. Shi-Li Zhang, Prof. Ulf Jansson and Ms. Iulia Brumboiu.